# Emergence, Construction, or Unlikely? Navigating the Space of Questions regarding Life's Origins

Stuart Bartlett[1,2,*] & Michael L Wong[3]
1) Division of Geological and Planetary Sciences,
   California Institute of Technology,
   1200 E California Blvd,
   Pasadena, CA 91125,
   United States.
   Email: sjbart@caltech.edu
2) Earth-Life Science Institute,
   Tokyo Institute of Technology,
   Tokyo,
   Japan
3) Earth & Planets Laboratory,
   Carnegie Institution for Science,
   5241 Broad Branch Rd NW,
   Washington, DC 20015,
   United States.
   Email: mwong@carnegiescience.edu

**Abstract:** We survey some of the philosophical challenges and pitfalls within origins research. Several of these challenges exhibit circularities, paradoxes, or anthropic biases. We present origins approaches in terms of three broad categories: unlikely (life's origin was a chance event), construction (life's origin was a stepwise series of synthesis and assembly processes), and emergence (life was always an amalgam of many parallel processes from which the living state emerged as a natural outcome of physical driving forces). We critically examine some of the founding and possibly misleading assumptions in these categories. Such assumptions need not be detrimental to scientific progress as long as their limits are respected. We conclude by attempting to concisely state the most significant enigmas still remaining in the origins field and suggest routes to solve them.

*Corresponding author: sjbart@caltech.edu

# How Can We Approach the Origins Quest(ion)?

What does it mean to research the origins of life (OoL)? It is fast becoming a scientific discipline in its own right, partly driven by its very fundamental philosophical appeal[1]. And, indeed, why should we not consider the origins of life a scientific problem? It seems clear that science offers the best chance of resolving the story of life. However, we should bear in mind some strong constraints on how it can be investigated. The most glaring issue is that we only have one tree of life[2] to study, and we cannot replay the tape or generate ensembles. This imbues us with a degree of prejudice, narrowing our focus to a small region of the space of possible living systems. Second, we cannot study the OoL as we can study traditional scientific problems, through repeated iterations of experimentation, observation, and theorizing. In fact, the OoL is somewhat closer to a forensic detective problem. However, the crime was committed a long time ago by such a sophisticated criminal that there is little to no evidence of the deed. In the spirit of detective investigations, we could turn to the great master himself, Mr. Sherlock Holmes, who suggested that "If you eliminate the impossible, whatever remains, however improbable, must be the truth." Eliminating the impossible is itself impossible for us because we cannot effectively explore the space of possible origins (being astronomical in size due to the combinatorics of, e.g., chemical reaction networks). Instead, we have to carefully examine the lasting consequences of the crime and retrodict backward to the source.

Much of modern origins research focuses on molecules, those that are deemed essential to life. The first issue with this perspective is that one could argue that life only uses essential molecules because there is a selection pressure against frivolous use of ancillary and "unnecessary" stuff (such stuff was perhaps useful in the past, but is now obsolete and, hence, absent). Animal counterexamples aside, we can, of course, point to highly simplified organisms, the prokaryotes, and "ancient" ones at that, and ponder their roots. They, of course, need genetic material, proteins, lipids, and sugars, among others. Hence, many origins researchers seek abiotic sources of amino acids, or the components of RNA, or lipids, or ideally all of the above. As a subproblem of organic synthesis, one can quite comfortably fill entire careers with this pursuit. But we should constantly remind ourselves where this pursuit might lead us. The day may well come (and soon) when we overcome this synthesis enigma and observe with some satisfaction a suite of organic molecules emerging from some kind of abiotic production line. But this molecular collection is not life, and if not prepared with a deeper understanding, we might be in for disappointment (see Construction section). We know from countless experimental endeavors that mixtures of organic molecules do not spring into life even when driven by various thermodynamic driving forces. We must have complementary theories and/or experiments that arm us with the knowledge of from where the dynamic organization of life comes, in particular the cryptic link between free energy dissipation and the emergence of information engines[3] and systems capable of learning.

---

[1] Not to mention the highly interdisciplinary nature of the problem and the likelihood of applied avenues that emerge en route.

[2] Our tree could be the result of multiple abiogeneses, and we also don't have access to the original abiogenesis event to study; we have only its ~4e9-year-old end products, plus a set of fossilized snapshots.

In this chapter, we take a frank look at the different approaches to the origins enigma in the context of similar scientific problems and some fundamental philosophical difficulties with the OoL. The interplay between chance and necessity is one of the most cryptic aspects of the biological world, and life and its history are full of surprises and frozen accidents that would not have been predictable a priori. Hence, we pose the following question: When the day comes that we "solve" the OoL, which news headline is more likely? "Scientists uncover origin of life: It was exactly what was predicted" or "Scientists uncover origin of life: Unconnected research stumbles upon completely unexpected answer."

## Avian Circularities

Some of the most prominent discussions in the origins field relate to the sequential order- ing of the various functions of life (chicken-and-egg problems). The most common is the metabolism/proteins/hardware vs. reproduction/nucleic acids/software debate. Researchers argue *ad nauseum* about which of these processes preceded the other. Proteins are nonfunctional unless they are accurately assembled with the appropriate amino acid sequence. But the genetic sequences that are read out by translation machinery cannot replicate themselves without highly sophisticated protein molecular machines. There are two other permutations in this debate: both metabolism and replication were always present (this idea is increasingly appearing in the literature), and neither came first (this possibility is rarely discussed). Recently, researchers investigating the structure and history of the ribosome have indeed written strongly in favor of the presence of both protein and nucleic acids from the earliest phases. This is motivated by the inextricable link between ribosomal RNA and ribosomal proteins in the function of the ribosome. Because it is a nexus of biological information processing, it is a natural starting point for seeking the origins of that information processing. Given the poor ability of proteins to store information, the similarly poor ability of nucleic acids to act as catalysts[4], and the highly likely possibility that prebiotic synthetic processes produce a broad range of biomolecular species (not just a narrow set), it seems most realistic to assume that the ancestors of proteins and nucleic acids existed alongside one another.

---

[3] Systems capable of making use of measurements or encoded information to convert one form of free energy into another. A prime example is protein-based molecular machines.

The fourth option, that neither protein nor nucleic acid was present in the beginning, should also be considered. Many alternative molecular systems for early life have been considered, including α-hydroxy acids in place of amino acids [5], thioesters in place of ATP [8] [9], and green rust as an information engine [27]. It seems plausible that the earliest life used different components to perform familiar tasks, albeit less efficiently; one does not expect a scaffolding to be identical to the finished product. The philosophical question left over with such an approach is how would a protolife system eventually transform and discover the biomolecules we are familiar with? The most frequent answer comes under the heading of chemical evolution. Despite this being a discipline in its own right, with its own journal, there have been relatively few genuine demonstrations of chemical evolution (note, however, the recent example of [29]), especially in the absence of nucleic acids. In particular, comprehensive experimental demonstration of the emergence of composomes within the GARD model would be highly compelling [19].

Overall, the view that nucleic acids or protein, reproduction or metabolism must come first and the other second is likely an artificial question. Any molecules that can some- how replicate have to have been synthesized somehow. That synthesis requires free energy, and hence, one could argue also a rudimentary metabolic process. On the other hand, any plausible prebiotic system that processes free energy and produces molecules with catalytic ability will probably exhibit some form of positive feedback or network autocatalysis [14] [31]. This type of feedback could be considered a basic form of reproduction because the action of a set of molecules has a positive causal effect on the further synthesis of those molecules. So the answer to the chicken-and-egg paradox is arguably becoming clear with modern advances: It is likely that both metabolism and reproduction were present from the outset. Perhaps they were also preceded by as yet unknown processes that were neither metabolism nor reproduction but, when viewed after the fact, were necessary stepping stones for life's foundations to emerge.

## Assuming that...

Science is designed to be axiomatic; a set of consequences and relations are derived from a basic set of assumptions. The history of science shows a consistent trend: At a given point in time, there is a set of assumptions and predictions that are tested and explored as far as possible. Eventually, scientific pursuits began to reach the limits of the knowledge status quo, and failures of existing theories began to accumulate. Some begin to question whether the basic assumptions might be too specific and overconstrained. With some combination of out-of-the-box thinking, imagination, creativity, experimentation, and luck, it is realized that a different, more general, or reduced set of assumptions are needed. This leads to a paradigm reset, an opening of new possibilities and new intellectual shores to be explored. In time, the limits of this new set of ideas will also be found. Whereas certainly providing considerable utility for a time, the new assumptions and theories will also eventually show their limitations, and those people seeking ultimate scientific truth will become dissatisfied and again consider disbanding the foundations on which that set of ideas are built. This is reminiscent of the idea of punctuated equilibrium: periods of incremental improvement, interspersed with dramatic jumps of innovation and change. For most of the history of origins research, a key assumption has been the association between life's identity and its molecular components: Life cannot exist in the absence of the nucleic acids, amino acids, lipids, and sugars that we see comprising life today (life likely used these molecules for most of its history). This assumption led to the quest for the prebiotic synthesis of biomolecules, which has had various highs and lows since the pioneering experiments of Urey and Miller. The assumption requires that any scenario for life's origins first explains the origins of life's known components and then explains the emergence of life's processes and functions. A potential corollary of this assumption is that the processes and functions of life are a natural and spontaneous consequence of the presence of biomolecules. But we have no guarantee nor any evidence of that. In the philosophical spirit, it is our duty to question our assumptions. Hence, is it only possible for life to originate with nucleic acids, amino acids, lipids, and sugars? If we deconstruct this assumption, we admit the possibility that the materials or components that are effective for *maintaining and evolving* life might be different from those that are effective for *starting* life.

---

[4] Though it appears that DNA is a highly versatile material that can perform all manner of structural, mechanical, and informational tasks. Note that many of these novel functions are artificially engineered and do not appear in extant life's repertoire [11] [15] [20] [28].

In many prebiotic chemistry experiments, there are significant quantities of "unwanted" synthesis products from side reactions. When such products are far from the "desired" molecular species, they are discarded as an unwanted distraction from the main show. Perhaps there should also be room for explorations of these less familiar species. In the spirit of novelty search [24], exploration of these molecules and systems may yield unexpected routes toward life. These routes may not follow our intuitions but instead follow the constraints of physics and the system boundary conditions. If those boundary conditions are aligned with prebiotic conditions, this approach would allow the systems themselves to tell us how life might have originated instead of following a storyline that we have come up with.

## Unlikely

The history of life is full of unpredictable, strange twists and turns, exaptations, and frozen accidents. Given the huge influence of happenstance and history in life's story and the fact that such stochasticity would have been at its strongest during life's beginnings, it is possible that a logical deconstruction or retrodiction of life's origins is impossible. Is the scientific method really able to answer the question of how life began?

Let us imagine that we know the answer, and it is that life began as a knock-on effect of an extremely unlikely fluctuation in a system that otherwise cares nothing for giving birth to life. This fluctuation was amplified by a series of downstream feedbacks that eventually led to life. If we were able to go back in time to damp out that fluctuation (as Q did with Picard in Star Trek: The Next Generation), the Earth would go on to remain sterile for its entire tenure.

The sad consequence of this reality is that we would be lonely in the universe, the sole winners of a game for which almost no planet becomes victorious. Science generally studies typical behavior, events that happen near the center of nicely shaped distributions. Heavy-tailed distributions, extreme events, and systems that depend heavily on their histories tend to be tricky to analyze using conventional statistics and experiments or simulations that are expected to be consistent and convergent. So if life's emergence is in the extreme end of an extreme distribution, does it render the scientific study of life's origins futile? Perhaps not.

Even if the OoL is unlikely in natural settings (making us quite lonely in the universe), one day we may be able to understand how to reliably create life artificially. Perhaps our experiments will discover ways of enhancing the "probabilities" of life's emergence by forcing the abiotic environment in one way or another (maybe there is a bottleneck in natural synthesis that we can overcome as we have done repeatedly in industrial projects, such as artificial nitrogen fixation to make fertilizer). Thus, in the space of all possibilities, there may be a corner in which the origin of life is both unlikely and scientifically explorable.

Indeed, we might find that the spontaneous emergence of life in the majority of planetary locales is close to impossible, but being inventive beings, we may discover our own improvised settings that, in fact, give rise to life readily. If this were true, despite our own origins being both unlikely and largely off limits to the approaches of science, a separate, synthetic origins discipline would emerge (indeed the field of artificial life already thrives) and with it the possible task of filling the sterile universe with life. One intriguing example of this is the idea of directed panspermia (also explored in *Star Trek: The Next Generation*), whereby intelligent life forms "seed" or otherwise induce the emergence of life on other worlds. In this case, the universe could be teeming with abiogenesis events, but the vast majority of them would have been generated by preexisting life.

Let us go one step further and imagine an even more pessimistic reality: that the OoL is both unlikely and not amenable to the scientific method, synthetic or otherwise. Is it possible to ever establish this as any kind of fact? Will we ever admit defeat? Say 200 years pass, and we still do not understand the emergence of life. Will we keep inventing new rabbit holes, each time thinking, "Ah, this one will finally lead us to the answer!" when in reality none can? Is there really a point at which the barren wasteland of OoL experiments will convince us that the OoL is not a scientific problem?

In fact, it probably does not matter. Given all the fascinating and useful discoveries en route to an understanding of the OoL, we will still benefit greatly from this pursuit even if its ultimate objective is destined to be eternally elusive. Thus, we need not lie awake at night worrying about the possibility that the OoL can never be understood. Even if it cannot, we can still do valuable and fulfilling work as we seek this illusory goal.

## Construction

*"Before life was life, it did not know what life was, nor that it was on its way to becoming life."*
—Murthy Gudipati (2018) [10]

Some storylines in origins research give the impression that, before life began, it knew where it was going and which steps would lead it there. This is a potentially misleading illusion that stems from viewing the problem through anthropic, engineer-like lenses. Pursuing the spontaneous synthesis of biomolecules goes all the way back to the founding experiments of Urey and Miller. Great progress since then, including analysis of meteoritic and cometary material, shows that amino acid synthesis is relatively common in a variety of scenarios [1] [18]. Nucleotide synthesis has proven to be dramatically less feasible with the most promising route at present relying on UV light and cyanide precursors [32]. Furthermore, large, unfolded polymers are extremely vulnerable to hydrolytic degradation, and difficult to ligate beyond a small number of monomeric units. Thus, the field focuses on the polymerization aspect intensely in recent decades, and these efforts have borne fruit in the form of thermal cycling mechanisms that can make polymerization favorable under certain conditions. Wet–dry cycles have become a popular method for defying the hydrolysis demon because it somewhat mimics the dehydration process that must accompany polymer growth [12] [22]. During the wet phase of the cycle, the requisite synthesis reactions produce monomers. As this system moves into the drying phase, evaporation removes water, making the mixture more concentrated and allowing the polymerization of the monomers to become more favorable. Rehydration causes a certain degree of hydrolytic damage, but the larger polymers can potentially resist through folding or self-shielding. Repeated cycles eventually produce longer molecules, perhaps up to ~100 monomers.

Note that other forms of thermodynamic cycling may also alleviate the polymerization problem. In particular, thermophoresis is shown to be effective at generating prebiotic polymers (e.g., [16]). In this process, convection cells in small, heated pores concentrate and polymerize due to thermal diffusivity effects as the molecules circulate between hot and cold regions.

Whether it is wet–dry cycles, thermophoresis, or mineral molecular machines, we will likely solve the polymerization riddle in the near future. Synthesis of all the key molecules of life as we know it will also probably follow in due course. And, alas, the origins researchers will finally rest as their work will have come to a grand conclusion. Or will they?

Let us imagine that at this advanced stage, the origins field has uncovered unequivocal methods for the synthesis and aggregation of all the key biomolecules: proteins, nucleic acids, lipids, sugars, and a few others for good measure. This appears to be the objective of a significant subset of the field, and let us envision the situation when this objective is entirely met. Will we have created life?

We can approach the problem from the opposite direction, an experiment and mental exercise that has been hotly debated for decades if not centuries: Can a sterilized sample of biomatter come to life? The short answer to this question is no, but as with all aspects of life, the reality is likely more complicated. When an experimentalist autoclaves a sample, that sample is then rendered nonliving by all standard definitions. Furthermore, there is no trivial or known way to reverse the sterilization process. What if the sample was *partially* autoclaved? By this, we mean what if the system was sterilized, but there were significant quantities of recognizable biomolecular species and structure remaining? Is there any way that *this* system could be physically or chemically manipulated such that it comes back to life? Perhaps, and indeed we suggest a quantitative version of this experiment to be carried out just in case the forces of self-organization might actually be capable of recapitulating life from a pot of broken life pieces. Conventional wisdom states that a pot of broken life pieces, the target of origins synthesis experiments, cannot come to life. What then will we do when the constructionist dream comes true? Even with all the molecules of life at hand, it is highly unlikely that we will have achieved a second genesis.

Most of physics is presented in the framework of driving forces and responses: An electromagnetic field compels the response of motion upon charged particles. Biology can similarly be cast in this light: competitive driving forces between species lead to the discovery of novel evolutionary innovations. Returning to the storyline above, if the constructionist dream comes true, we will have discovered driving forces in the form of precursor supplies and free energy sources, and responses in the form of the desired biomolecules. There is a large missing piece here though, namely, what are the driving forces that lead to the response of the formation of a living organism? Just the *presence* of molecular components is likely an insufficient driving force for the emergence of life. So, at this point, the question becomes what class of boundary conditions might compel a system rich in biomolecules to form a protocell or progenote? Indeed, at this point, it could be that the constructionist narrative reduces to the unlikely narrative: if making the molecules of life is straightforward but pushing them to form life itself is a next to impossible feat of extreme events.

There is, of course, a more optimistic possibility: that in finding out how to make the material of life, the conditions for that material to form life naturally suggest themselves. In other words, perhaps the emergence of life was a by-product of what was initially an act of chemical synthesis. Perhaps the specific chemical routes to biomolecules are somehow predestined to produce the processes of life as well as the material of life.

Whereas this may very well be true, overzealous hopefulness that function follows form may blind us to lifelike functionality that exists in different forms. An alternative school of thought considers the processes of life, irrespective of its components. To illustrate this point, we present the following analogy (also given in [3]). Figure 1 shows a JR Central SCMaglev Shinkansen.



**Fig. 1**: A JR Central Maglev Shinkansen during performance tests. *Image due to Hisagi, Wikimedia Commons.*

This is one of the most sophisticated locomotives ever built with a top speed of 603 km/h. It uses magnetic suspension and levitates approximately 10 cm off the ground. Imagine that distant aliens are bored and fancy a new game to play. They open a wormhole to Earth and steal one of these trains. The train is given to some friends who know nothing about Earth and humans. The game is simple: figure out what the first ever locomotive was like. Direct observation of the Shinkansen would reveal the various materials: carbon fiber, titanium, aluminum, silicon, etc. The control system uses silicon-based electronics, logic circuits, and perhaps a neural network or two. A not-so-clever alien might conclude that the first locomotive was built from less sophisticated artificial alloys and was run by a primitive electronic computer (both of which would be incorrect). A more clever alien would realize that the original locomotive would have been built in a time before Maglev technology and, therefore, skated along oiled surfaces or tracks made of ice (which would also be incorrect). An even cleverer alien might liberate their thoughts from the restrictive morphology of the Shinkansen and invoke alternative components, such as wheels and an engine that runs on liquid propellant (closer, but still not entirely correct). The cleverest alien would spend more time contemplating the *purpose* of the locomotive, remain skeptical about the prospects of knowing exactly the original train's components, and simply deduce that ancient locomotives needed a source of electrons and dissipated those electrons as heat via the generation of a motive force.

**Emergence**

On the other end of the chance–determinism scale, we have emergence. This paradigm might suggest that life was somehow destined to begin on Earth for reasons we do not yet understand, but are likely related to basic physical tendencies, such as the second law of thermodynamics and the principle of least action. As a strongly driven, open system, the Earth is teeming with gradients and free energy sources of myriad forms, and this has been true throughout its entire story.

Emergence refers to the appearance of novel structures or processes in a system that either did not exhibit such features previously or did not exhibit them at a lower level (such as a smaller length scale or a smaller population of particles). Arguably the best understood example is the emergence of the ordered ferromagnetic phase in the Ising model of spin systems. The Ising model consists of a lattice of magnetic dipoles that interact with one another locally and are also influenced by a background magnetic field. Above the so-called critical temperature, the spins are completely uncorrelated due to the disordering effects of thermal noise. However, when the temperature passes below the critical temperature, a dramatic global shift occurs. Islands of correlated spins appear and grow with one domain eventually growing to the length scale of the whole system. This is the ferromagnetic phase. One of the most interesting aspects of this simple system is its behavior when the temperature is held at or infinitesimally close to the critical temperature. The name "critical" is used because, at this point, the system cannot decide whether to be in the paramagnetic (disordered) or ferromagnetic phase. As a result, it exhibits properties of both with ordered domains of finite sizes, but none dominating the whole system. In fact, the distribution of domain sizes is scale free, so the system also does not decide on a preferred length scale.

The appearance of fractal domains at the critical temperature is another example of emergence. There is nothing in the Hamiltonian of the Ising model that relates to scale-free domains; this feature is emergent. However, another virtue of the Ising model is that it is exactly solvable and is even tractable in two dimensions (though not at higher dimensionalities). So this emergent feature is calculable. The mathematical approach that allows the prediction of emergent properties is known as renormalization group theory. Whereas this is a powerful technique, it is not straightforward to apply to many systems of interest.

For this reason, emergence often remains shrouded in mystery. We associate emergence with nonlinear and networked systems, and often the emergent features can be described mathematically but not derived from the equations of motion of the system.

Fluid turbulence is another classic example. There is no known violation of the Navier– Stokes equations for fluid flows; hence, we understand the lowest level of continuous fluid motion. Turbulent flows exhibit an immediately recognizable and deep level of structure (see Figure 2 for an example). Both time series and spatial fields from fluid flows show patterns that can be readily described mathematically. However, no one has succeeded at a universal theory for turbulence.

Emergence plays a key role in our understanding of the living world. Alongside the origins of life, a related and similarly perplexing problem is the emergence of consciousness and thought patterns from the dynamics of neurons and synapses. It remains to be seen whether it is even possible to derive the higher level processes of the brain from the dynamics of its components (depending on the role of top-down causation and other compounding effects), but recent advances suggest that one day this goal might be achieved [21].

The deepest theme that permeates emergence questions such as these is causality: Why do nonequilibrium systems tend to form these complex patterns? Authors in the field note that all dynamical systems compute and process information [36] [37]. Of course, many dynamical systems compute in trivial ways: They either maintain the information through time (frozen systems with little or no dynamics), or they erase and generate information at equal rates (completely random systems or isolated systems at equilibrium). However, the natural world is also full of examples that combine these two extremes, and the biosphere is a case in point. It is capable of storing information over billions of years if not longer and also exhibits chaos. Furthermore, it can exploit encoded information in extraordinarily sophisticated ways, the extent of which is still incompletely explored.

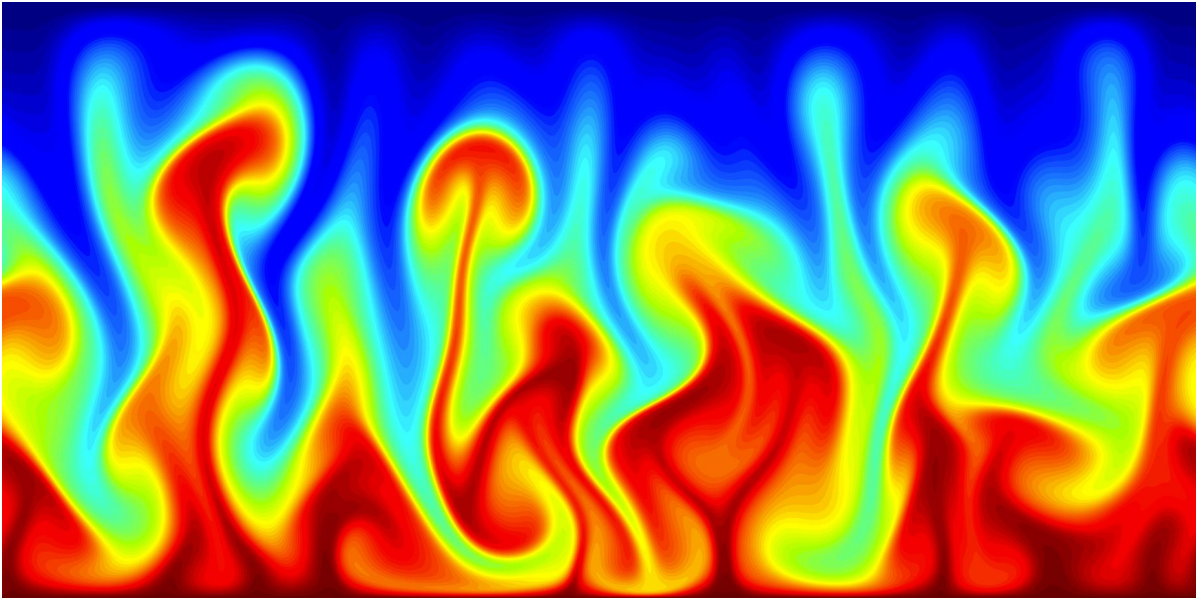

**Fig 2.** A turbulent flow field in a differentially heated fluid. Although bulk and statistical properties of such systems can be calculated, there is no universal theory that can predict the exact flow dynamics of such systems.

For the simpler examples of emergence, their causal relations can be traced back to fundamental physical principles. In the case of the Ising model, the ferromagnetic phase emerges from energy minimization. The behavior of the ideal gas can also be traced back to the kinetic behavior of its constituent particles. Moving toward the origins of life, driven chemical systems synthesize particular molecules due to a combination of thermodynamic and kinetic effects: Reactions bring a system toward the equilibrium state, subject to the constraints of energetic barriers to transition states. Even if someone were to achieve the constructionist dream of producing all the desired molecules in an integrated system, we may be able to map the entire reaction network and understand its steady states and kinetics. But what about the processes of life because we argued previously that the material of life is insufficient to constitute life itself? In our opinion, the most comprehensive theory for life would elucidate the causal relations between physical driving forces and the emergence of information processing that distinguishes life from nonlife. For many, this means understanding how and why the ribosome and the genetic code emerged [4] [30]. For others, this might mean a more basic emergence of learning behavior in physicochemical systems, such as an enzymatic reaction network capable of associative learning [13] [23] [33].

At its core, life is a learning system fueled by free energy sources. We see this at every level of its hierarchy from single molecules to the digital realm that connects human minds across the entire planet. Yet do we understand how and why a single molecule might process information? Significant advances in molecular biology, information thermodynamics, and stochastic thermodynamics reveal in quantitative detail how biological information engines can convert between forms of energy and information. Hence, the how and why of biomolecular information processing in extant life is gradually becoming understood. What remains a completely open question is how and why a nonequilibrium system with a given composition might produce emergent structures that *spontaneously* process information. Whereas real Maxwell demons at the scale of molecules to single electrons have been examined in laboratory settings [17] [25] [26], we still do not understand how natural information engines first arose.

In closing this chapter, we speculate as to the general conditions that might need to be satisfied for such a transition to occur. At a minimum, the system in question would need to contain molecular components that can definitely form information-processing structures (peptides and nucleic acids are the obvious candidates, but other molecules may be better suited; this remains to be seen). Second, there would need to be signals or processes within the system or on its boundary that are learnable [2]. In other words, a system that is driven in a trivial way is unlikely to produce emergent learning systems because there has to be something to learn or a problem to solve [34]. Third, there needs to be a way for the candidate learning structures to interact with the learnable signals and with one another (communication channels and rudimentary sensors). Finally, there needs to be some kind of stabilizing feedback, such that systems exhibiting learning behavior are more dynamically stable as a result of their learning behavior. When these conditions are satisfied, we might witness emergent learners spontaneously learning, possibly en route to the living state.

## References

1. Altwegg, K., Balsiger, H., Bar-Nun, A., Berthelier, J.J., Bieler, A., Bochsler, P., De Keyser, J., Prebiotic chemicals—amino acid and phosphorus—in the coma of comet 67P/Churyumov- Gerasimenko. Sci. Adv., 2, 5, #e1600285, 2016.
2. Bartlett, S.J. and Beckett, P., Probing complexity: Thermodynamics and computational mechanics approaches to origins studies. Interface Focus, 9, 6, #20190058, 2019.

3. Bartlett, S. and Wong, M.L., Defining Lyfe in the Universe: From Three Privileged Functions to Four Pillars. Life, 10, 4, 42, 2020.
4. Bernier, C.R., Petrov, A.S., Kovacs, N.A., Penev, P.I., Williams, L.D., Translation: The univer- sal structural core of life. Mol. Biol. Evol., 35, 8, 2065–2076, 2018.
5. Chandru, K., Mamajanov, I., Cleaves, H.J., Jia, T.Z., Polyesters as a model system for building primitive biologies from non-biological prebiotic chemistry. Life, 10, 1, 6, 2020.
6. Crutchfield, J.P., Between order and chaos. Nat. Phys., 8, 1, 17–24, 2012.
7. Feldman, D.P., McTague, C.S., Crutchfield, J.P., The organization of intrinsic computation: Complexity-entropy diagrams and the diversity of natural information processin. Chaos: Interdiscip. J. Nonlinear Sci., 18, 4, #043106, 2008.
8. Goldford, J.E., Hartman, H., Smith, T.F., Segre, D., Remnants of an ancient metabolism without phosphate. Cell, 168, 6, 1126–1134, 2017.
9. Goldford, J.E., Hartman, H., Marsland, R., Segre, D., Environmental boundary conditions for the origin of life converge to an organo-sulfur metabolism. Nat. Ecol. Evol., 3, 12, 1715–1724, 2019.
10. Gudipati, M., Personal communication, 2018.
11. Hamada, S., Yancey, K.G., Pardo, Y., Gan, M., Vanatta, M., An, D., Hu, Y. et al., Dynamic DNA material with emergent locomotion behavior powered by artificial metabolism. Sci. Robot., 4, 29, #eaaw3512, 2019.
12. Higgs, P.G., The effect of limited diffusion and wet-dry cycling on reversible polymerization reactions: implications for prebiotic synthesis of nucleic acids. Life, 6, 2, 24, 2016.
13. Hjelmfelt, A., Weinberger, E.D., Ross, J., Chemical implementation of neural networks and Turing machines. Proc. Natl. Acad. Sci., 88, 10983–10987, 1991.
14. Hordijk, W. and Steel, M., Autocatalytic networks at the basis of life's origin and organization. Life, 8, 4, #62, 2018.
15. Joesaar, A., Yang, S., Bogels, B., van der Linden, A., Pieters, P., Kumar, B. V. V. S., Dalchau, N., Phillips, A., Mann, S., de Greef, TF., DNA-based communication in populations of synthetic protocells. Nat. Nanotechnol., 14, 4, 369–378, 2019.
16. Kreysing, M., Keil, L., Lanzmich, S., Braun, D., Heat flux across an open pore enables the continuous replication and selection of oligonucleotides towards increasing length. Nat. Chem., 7, 3, 203–208, 2015.
17. Kutvonen, A., Koski, J., Ala-Nissila, T., Thermodynamics and efficiency of an autonomous on-chip Maxwell's demon. Sci. Rep., 6, #21126, 2016.
18. Kvenvolden, K., Lawless, J., Pering, K., Peterson, E., Flores, J., Ponnamperuma, C., Moore, C., Evidence for extraterrestrial amino-acids and hydrocarbons in the Murchison meteorite. Nature, 228, 5275, 923–926, 1970.

19. Lancet, D., Zidovetzki, R., Markovitch, O., Systems protobiology: Origin of life in lipid catalytic networks. J. R. Soc. Interface, 15, 144, #20180159, 2018.
20. List, J., Falgenhauer, E., Kopperger, E., Pardatscher, G., Simmel, F.C., Long-range movement of large mechanically interlocked DNA nanostructures. Nat. Commun., 7, 1, 1–7, 2016.
21. Makin, J.G., Moses, D.A., Chang, E.F., Machine translation of cortical activity to text with an encoder–decoder framework. Nat. Neurosci., 23, 4, 575–582, 2020.
22. Mamajanov, I., MacDonald, P. J., Ying, J., Duncanson, D. M., Dowdy, G. R., Walker, C. A., Engelhart, A. E., et al., Ester formation and hydrolysis during wet-dry cycles: Generation of far-from-equilibrium polymers in a model prebiotic reaction. Macromolecules, 47, 4, 1334–1343, 2014.
23. McGregor, S., Vasas, V., Husbands, P., Fernando, C., Evolution of associative learning in chemical networks. PloS Comput. Biol., 8, 11, #e1002739, 2012.
24. Lehman, J. and Stanley, K.O., Why Greatness Cannot be Planned: The Myth of the Objective, Springer International Publishing, Germany, 2015.
25. Naghiloo, M., Alonso, J.J., Romito, A., Lutz, E., Murch, K.W., Information gain and loss for a quantum maxwell's demon. Phys. Rev. Lett., 121, 3, #030604, 2018.
26. Ribezzi-Crivellari, M. and Ritort, F., Large work extraction and the Landauer limit in a con- tinuous Maxwell demon. Nat. Phys., 15, 7, 660–664, 2019.
27. Russell, M.J., Green rust: the simple organizing 'seed' of all life? Life, 8, 3, 35, 2018.
28. Thubagere, A.J., Li, W., Johnson, R.F., Chen, Z., Doroudi, S., Lee, Y.L., Winfree, E., Izatt, G., et al., A cargo-sorting DNA robot. Science, 357, 6356, #eaan6558, 2017.
29. Vincent, L., Berg, M., Krismer, M., Saghafi, S.T., Cosby, J., Sankari, T., Baum, D.A., Chemical ecosystem selection on mineral surfaces reveals long-term dynamics consistent with the spontaneous emergence of mutual catalysis. Life, 9, 4, 80, 2019.
30. Vitas, M. and Dobovisek, A., In the beginning was a mutualism-on the origin of translation. Orig. Life Evol. Biosph., 48, 2, 223–243, 2018.
31. Xavier, J.C., Hordijk, W., Kauffman, S., Steel, M., Martin, W.F., Autocatalytic chemical networks at the origin of metabolism. Proc. Royal Soc. B, 287, #20192377, 2020.
32. Xu, J., Chmela, V., Green, N.J., Russell, D.A., Janicki, M.J., Góra, R.W., Szabla, R., Bond, A.D., Sutherland, J.D., Selective prebiotic formation of RNA pyrimidine and DNA purine nucleo-sides. Nature, 582, 7810, 60–66, 2020.
33. Poole, W., Ortiz-Munoz, A., Behera, A., Jones, N. S., Ouldridge, T. E., Winfree, E., Gopalkrishnan, M., Chemical boltzmann machines, in: International

Conference on DNA Based Computers, pp. 210–231, Springer, Cham, 2017, September.
34. Wong, M. L., Bartlett, S., Chen, S., Tierney, L., Searching for life, mindful of lyfe's possibilities. Life, 12, 6, #783, 2022.